\documentstyle[12pt]{article}
\input amssym.def 
\input amssym 
\newcommand{\Bbc}[1]{{\Bbb{#1}}} 
\newcommand{\Real}{\Bbc{R}}
\newcommand{\Dif}{{\cal{D}}}
\newcommand{\Str}{\rm{Supertrace}}
\newcommand{\Tr}{\rm{Trace}}
\newcommand{\Comm}{[\Omega,\chi]}
\newcommand{\End}{\nonumber\\}
\begin{document}
\begin{flushright}
   \large{KCL/MTH-97-01\\
   hep-th/9701070}
\end{flushright}
\begin{center}
\Huge{The Gauge-Fixing Fermion in BRST Quantisation}\\
\ \\
\large{Lecture given at the conference} \\
\Large{Constrained Dynamics and Quantum Gravity II\\
  Santa Margherita, Italy, September 1996 }\\
\ \\
\Large{Alice Rogers\\
       Department of Mathematics \\
        King's College \\
        Strand\\
         London WC2R 2LS, Great Britain}\\
\ \\
\Large{January 9, 1997}        
  \\
\end{center}

\begin{abstract}
Conditions which must be satisfied by the gauge-fixing fermion $\chi$ 
used in the BRST quantisation of constrained systems are established. 
These ensure that the extension of the Hamiltonian by the gauge-fixing 
term $\Comm$ (where $\Omega$ is the BRST charge) gives the correct path 
integral.
\end{abstract}
%
In canonical BRST quantisation (particularly the path-integral approach 
developed 
by Batalin, Fradkin, Fradkina and Vilkovisky in a series of 
papers \cite{fraVil1,BatVil,FraFra,BatFra,FraVil2} (BFV)  and Henneaux 
\cite{Hennea}) the starting point is the simple 
$2n$-dimensional phase space $\Real^{2n}$ with $n$ position 
coordinates $q^i$
and $n$ momenta $p_i$, together with $m$ first class constraints 
$T_a(p,q) = 0$ and a first class Hamiltonian $H(p,q)$. (Here and later, 
the integer $n$ is greater than $m$, the index $i$ runs from $1$ to $n$ 
and the index $a$ runs from $1$ to $m$.) The presence of 
the constraints means that the true phase space has the more complicated 
structure of a quotient space of a non-trivial manifold; the key idea in 
BRST quantization is that cohomology classes on a simpler extended phase 
space are used 
instead of observables on this complicated true phase space. 

The 
extended phase space used is the $(2n+2m,4m)$-dimensional superspace 
$\Real^{2n+2m,4m}$ with coordinates $p_i, q^i, k_a, l^a, \eta^a, 
\pi_a, \theta^a, \phi_a $, where the $k_a$ and $l^a$ are commuting 
conjugate 
pairs, while the $\eta^a$ and $\pi_a$ are anticommuting conjugate pairs, 
as are the $\theta^a$ and the $\phi_a$. (The anticommuting coordinates 
correspond to ghosts, antighosts and their momenta. Ghost number is 
defined so that $\eta^a$ and $\theta^a$ have ghost number $1$ while 
$\pi_a$ and $\phi_a$ have ghost number $-1$.) 
Collectively these 
coordinates are denoted $P,Q$. The BRST charge $\Omega$, which has ghost 
number $1$, has the form
\begin{equation}
\Omega = T_a \eta^a + k_a \theta^a + {\rm\ higher\ order\ terms},
\end{equation}
the terms of higher order in the anticommuting variables being 
determined by the requirement that $\Omega^2= 0$. (Quantisation is 
carried out by the standard canonical procedure, with states being 
functions 
$f(q,l,\eta,\phi)$ and the corresponding conjugate momenta $p,k,\pi$ and 
$\theta$ defined as derivatives in the normal manner.)

The main result given by BFV and Henneaux is that the vacuum 
expectation value for the theory has the path integral expression
\begin{equation}
\int \Dif P \Dif Q 
 \exp i\Bigl(  \int_0^t  p_a \dot{q}^a + k_a \dot{l}^a  
 \, + \pi_a \dot{\eta}^a + \theta^a \dot{\phi}_a 
+ H(P,Q)+ K(P,Q)\,dt \Bigr)
\label{PI} \end{equation}
where $K= \Comm$ is the commutator of the BRST charge $\Omega$ with a 
field $\chi$ of ghost number $-1$ (the 
gauge-fixing fermion), and the path integral is taken over closed paths. 
(In 
some cases $H$ must be extended by terms involving ghosts so that the 
modified Hamiltonian commutes with the BRST charge $\Omega$.)

In the work of Henneaux it is implied that the gauge-fixing fermion 
$\chi$ is 
arbitrary; however, it is also clear that $\chi$ cannot, for instance, 
be zero, and so one sees that some 
non-singularity condition is called for. In the BFV papers, the 
gauge-fixing fermion is assumed to take the form
\begin{equation}
\chi = l^a \pi_a + X^a \phi_a + {\rm\ higher\ order\ terms}
\label{BFVXI}\end{equation}
where the $m$ even functions $X^a$ must be such that the matrix formed 
by the commutators $[T_a,X^b]$ is non-singular. (In fact the proof given 
by BFV is only valid when the constraints commute, but the more general 
result has recently been 
established by Batalin and Marnelius \cite{BatMar}.) 

The purpose of this talk is to describe a criterion which the 
gauge-fixing fermion must satisfy, together with a simple proof of the 
main result. This is first to clarify existing work and secondly to 
assist the study of
cases where the Gribov problem might appear to prevent the 
existence of a gauge-fixing fermion.

The starting point is the knowledge that the true space of states is 
$H_0(\Omega)$, the $\Omega$-cohomology group of states of ghost number 
zero.  The first observation is that if $L$ is an operator which 
commutes with $\Omega$ and has ghost number zero,  then (as shown by 
Schwarz \cite{Schwar})
\begin{equation}
\Str\, L = \sum_{k=-m}^{k= +m} (-1)^k \Tr_{H_k(\Omega)} L. 
\end{equation}
The only contribution to the supertrace comes from $\Omega$-closed 
states which are not exact, since, if $L f = \lambda f$ then $L \Omega f 
= \lambda \Omega f$, so 
that if $\Omega f $ is not zero we have two states with ghost number 
differing by one and equal $L$ eigenvalues; thus the contribution from 
non-closed states cancels with that from exact states. (This resembles 
the argument used by 
McKean and Singer in the context of index theory \cite{McKSin})

A second observation is that the path integral (\ref{PI}), being over 
closed paths, gives the supertrace of $\exp i(H+ \Comm)t$ -- provided of 
course that this operator does have a 
supertrace -- so that, if all cohomology groups other 
than the zeroth one vanish, the superspace path integral reduces to a 
trace over the space $H_0(\Omega)$ of physical states. The path integral 
(\ref{PI}) will then give the required physical result.

The gauge-fixing fermion must thus be chosen so that the commutator 
$\Comm$ can play a crucial double role, both ensuring the vanishing of 
the cohomolgy groups other than that at zero ghost number, and making 
the extended Hamiltonian sufficiently regular for the operator $\exp 
i(H+ \Comm)t$ to have a well-defined supertrace. 
We start with the simple fact that if $\Comm$ were an invertible 
operator we would have no $\Omega$-cohomology at all. This may be shown 
by considering a state $f$ such that $\Omega f = 0$; then, if $h= \Comm 
f$, 
\begin{eqnarray}
f = \Comm^{-1} h &=& \Comm \Comm^{-2} h \End
   &=& (\Omega \chi + \chi \Omega) \Comm^{-2}h \End
   &=& \Omega \chi \Comm^{-2} h + \chi \Comm^{-2} \Omega f \End
   &=& \Omega \chi \Comm^{-2} h,
\end{eqnarray}
so that $f$ must be cohomologically trivial. (Here the fact that 
$\Omega$ commutes with $\Comm$, and hence with $\Comm^{-1}$ has been 
used.)

Now of course we do want some cohomology; if $\chi$ existed such that 
$\Comm$ was invertible we would have no physical states; what in fact we 
require is that $\chi$ be such that {\em the only states on which 
$\Comm$ 
is not invertible are zero ghost number states which are not exact}. 
This then ensures that there is no cohomology except at ghost number 
zero. This criterion is not quite sufficient to ensure that the path 
integral (\ref{PI}) gives the correct result, because the operator $H = 
\Comm$ may not be sufficiently regular to have a well-defined 
supertrace. The additional condition which $\Comm$ must satisfy is that 
$H + \Comm$ {\em should have positive discrete eigenvalues tending to 
infinity in each ghost sector}, ensuring the necessary absolute 
convergence of the sums involved in the supertrace.

In the standard simple example it may easily be seen that these 
conditions are satisfied. In this example, where $T_a = p_a$ and $X^a = 
q^a$, the quartet mechanism of Henneaux and Teitelboim \cite{HenTei} 
shows that $\Comm$ is a number operator counting ghost and gauge number, 
and has precisely the required analytic properties.  

A longer paper, 
with more details of this work together with non-trivial examples, is in 
preparation. Given the group theoretic analysis of BRST cohomology 
by Kostant and Sternberg \cite{KosSte}, it is possible that a 
development of recent 
interesting work of Klauder \cite{Klaude} using coherent states can be 
used to provide a general method for applying the BFV approach to 
systems affected by the Gribov problem.  The relationship of the 
construction of the gauge-fixing fermion used by McMullan \cite{McMull} 
to the analysis in this paper would also be interesting.

\end{document}